# Universal heat transport in a heavy-fermion superconductor

H. Shakeripour [1], M.A. Tanatar [1] [*], C. Petrovic [2,3] & Louis Taillefer [1,3] [†]

*1 Département de physique and RQMP, Université de Sherbrooke, Sherbrooke, Québec J1K 2R1, Canada*

*2 Condensed Matter Physics and Materials Science Department, Brookhaven National Laboratory, Upton, New York 11973, USA*

*3 Canadian Institute for Advanced Research, Toronto, Ontario M5G 1Z8, Canada*

[*] Present address: Ames Laboratory, Ames, Iowa 50011 USA

[†] e-mail: louis.taillefer@physique.usherbrooke.ca.

**The symmetry of the order parameter is the defining property of a superconductor and a strong clue to its pairing mechanism[1]. While *d*-wave symmetry is firmly established in high-$T_c$ superconductors[2], in no heavy-fermion superconductor is the symmetry known definitively. One way to elucidate the order parameter is to locate the symmetry-imposed nodes in the energy gap by measuring the heat carried by the zero-energy quasiparticles associated with those nodes as a function of direction. As verified in high-$T_c$ superconductors[3,4], a line of nodes yields universal heat transport[5,6], independent of impurity scattering. Here we show that heat conduction in the heavy-fermion superconductor CeIrIn$_5$ is also universal, but only for a current perpendicular to the tetragonal axis of its crystal structure, not parallel to it. This uniaxial anisotropy is strong evidence for a line of nodes in the basal plane, which is inconsistent with the *d*-wave symmetry proposed for the isostructural superconductor CeCoIn$_5$ (refs. 7, 8, 9). Different symmetries in the two materials would explain why the phase diagram of these heavy-fermion compounds consists of two separate superconducting domes[10].**

In certain symmetries, the energy gap of a superconductor vanishes along a line on the Fermi surface. For example, in the $B_{1g}$ representation of the tetragonal $D_{4h}$ point group, the $d_{x^2-y^2}$ state transforms as $k_x^2 - k_y^2$ so that the line nodes are located where the $k_x = k_y$ and $k_x = -k_y$ planes intersect the Fermi surface[2]. A line node produces a density of states which grows linearly with energy at low energy. Universal heat conduction is a direct consequence of such a linear density of states, the result of a compensation between the growth in the zero-energy quasiparticle density of states and the decrease in the quasiparticle mean free path with increasing impurity scattering[5,6]. In the $T = 0$ limit, the electronic thermal conductivity divided by temperature, $\kappa / T$, extrapolates to a finite residual linear term, $\kappa_0 / T$, which is independent of impurity scattering and governed only by the quasiparticle velocities normal and tangential to the Fermi surface at the node[5,6]. Experimentally confirmed in cuprates[3,4] and in ruthenates[11], universal heat conduction has not been observed in a heavy-fermion superconductor until now[12,13], even though in most cases a line node in the gap has been invoked to account for the properties of their superconducting state.

Here we report on a study of heat transport in the heavy-fermion superconductor $CeIrIn_5$, whose critical temperature $T_c \approx 0.4$ K (ref. 14). In Fig. 1, the thermal conductivity of nominally pure samples of $CeIrIn_5$ is compared to that of samples doped with 0.1% La ($Ce_{0.999}La_{0.001}IrIn_5$), for a heat current parallel ($J \parallel c$) and perpendicular ($J \parallel a$) to the $c$-axis of the tetragonal structure. Data for the pure samples were reported in a previous publication[15], whose main finding was a pronounced $a$-$c$ anisotropy, highlighted in Fig. 2. While $\kappa_a(T) / T$ extrapolates to a sizable residual linear term, $\kappa_{0a} / T \approx 20$ mW / K$^2$ cm, in quantitative agreement with theoretical expectation for a line node[5,15], $\kappa_c(T) / T$ extrapolates to a negligible value[15].

As seen in Fig. 1a, $\kappa_a / T$ in the pure sample is close to linear all the way from slightly below $T_c$ down to the lowest measured temperature. La doping does not alter this dependence at high temperature, but it causes upward curvature in $\kappa_a/T$ at low temperature (see Fig. 1a), with a $T^2$ behaviour below 0.15 K. The development of curvature at low temperature is theoretically expected for a line node, at temperatures below the impurity bandwidth $\gamma$ (ref. 5). The fact that no curvature is observed in the pure sample is consistent with the smaller $\gamma$ in that sample. The two curves are seen to come together in the $T \to 0$ limit, revealing that conduction in this direction is universal. We conclude that the superconducting gap of CeIrIn$_5$ definitely contains a line node. This is consistent with all other properties of this material, including the specific heat[16] ($C_e \sim T^2$) and the NMR relaxation rate[17] ($T_1^{-1} \sim T^3$).

Let us now turn to the other current direction, along the $c$-axis. In the pure sample, $\kappa_c/T$ shows a downward curvature at low temperature such that a linear extrapolation to $T = 0$ yields a negative residual linear term, implying that $\kappa_c/T$ must acquire upward curvature below our lowest data point, as sketched by the red line in Fig. 1b. La doping results in the unambiguous appearance of a sizable residual linear term $\kappa_{0c}/T$. This shows that heat conduction along the $c$-axis is not universal.

In Fig. 3, the value of $\kappa_0/T$ for several samples is plotted as a function of their residual resistivity $\rho_0$. While the large $\kappa_0/T$ along the $a$-axis is independent of $\rho_0$, $\kappa_0/T$ along the $c$-axis is proportional to $\rho_0$, vanishing in the limit of $\rho_0 \to 0$. This is precisely what is expected for a gap that has a line node in the basal plane. In $D_{4h}$ symmetry, two states have such a line node: 1) the $k_z(k_x + i k_y)$ state in the even-parity $E_g$ representation[5,18], whose "hybrid gap" – drawn in Fig. 2b – also has linear point nodes along the $c$-axis (*i.e.* point nodes at which the gap vanishes linearly with angle[5]); 2) the $k_z$ state in the odd-parity $A_{2u}$ representation (in weak spin-orbit coupling)[18], whose "polar gap" – drawn in Fig. 2c – has no other nodes. In both cases, $\kappa_{0c}/T$ is predicted to grow as a function of $\gamma$, linearly for the hybrid gap, quadratically for the polar gap[5,18]. The dependence of $\gamma$ on the normal-state impurity scattering rate $\Gamma_0$ depends on the scattering phase shift; in the strong limit of resonant scattering ($\pi/2$ phase shift), $\gamma \sim (\Delta_0 \Gamma_0)^{1/2}$ (ref. 5). Assuming that impurity scattering can be described by a single scattering rate $\Gamma_0$, we have $\Gamma_0 \sim \rho_0$, and in that limit $\kappa_{0c} / T \sim \gamma \sim (\rho_0)^{1/2}$ for a hybrid gap[5] and $\kappa_{0c} / T \sim \gamma^2 \sim \rho_0$ for a polar gap[18]. This would give a dependence of $\kappa_{0c} / T$ on $\rho_0$ as indicated by the dashed and solid lines in Fig. 3, respectively. Within error bars, both dependences are compatible with the data.

The pronounced and qualitative uniaxial anisotropy revealed by heat transport in CeIrIn$_5$ is inconsistent with $d$-wave symmetry, whether it be the $d_{x^2-y^2}$ state of the $B_{1g}$ representation (which transforms as $k_x^2 - k_y^2$) or the $d_{xy}$ state in the $B_{2g}$ representation (which transforms as $k_x k_y$)[2,18]. In such states, the gap would go to zero where planes containing the $c$-axis ($k_x = \pm k_y$ or $k_x, k_y = 0$) cut the Fermi surface. The resulting "vertical" line nodes would make quasiparticle conduction along $a$ and $c$ fundamentally similar. It has been shown[19] that even within $d$-wave symmetry some uniaxial anisotropy can be generated by requiring that the $c$-axis dispersion of one Fermi surface sheet, the quasi-2D open cylinder[20,21], be accidentally small along the nodal directions. However, this artificial constraint would not in general apply to or be effective on the other 4 or 5 sheets of the complex Fermi surface of CeIrIn$_5$ (refs. 20, 21). By contrast, any order parameter in the $D_{4h}$ point group that requires the superconducting gap to vanish in the $k_z = 0$ plane by symmetry will automatically produce the qualitative anisotropy seen here[5,18,19], for all Fermi surface sheets, whatever their shape, dispersion and location.

In a recent study of heat transport as a function of magnetic field direction in the basal plane, a tiny four-fold variation of $\kappa$ vs field angle, of magnitude 1 % of the normal-state conductivity, was interpreted as evidence for a $d$-wave gap[22]. The much larger uniaxial anisotropy discussed here, of magnitude 300 % of the normal-state anisotropy at $T = T_c / 6$, and increasing as $T \to 0$ (see Fig. 2), indicates that this four-fold variation does not reflect the dominant nodal structure. It could instead come from an in-plane anisotropy of the effective mass on whatever Fermi surface dominates the heat transport at a particular field. In a multi-band superconductor like CeIrIn$_5$, this dominance will shift with field from one Fermi surface sheet to another if the gap is band dependent, as is dramatically the case in CeCoIn$_5$ (ref. 13). This could explain why the four-fold variation changes phase with field[22].

We now compare CeIrIn$_5$ with CeCoIn$_5$, whose critical temperature $T_c \approx 2.4$ K. In CeCoIn$_5$, a large residual linear term $\kappa_0 / T$ was found for transport in both directions[13]. With the addition of $x$ La impurities per Ce atom, $\kappa_0 / T$ was seen to drop rapidly (for both directions), tracking the suppression of the normal-state conductivity $\kappa_N / T \sim 1 / x \sim 1 / \rho_0$ (ref. 13). This is in marked contrast to the behaviour of the specific heat, which showed an increase of its residual linear term with doping[13], as expected of a nodal superconductor. The difference was interpreted in terms of an extreme form of multi-band superconductivity, whereby one sheet of the Fermi surface – of high mobility and light mass – has a gap of negligible magnitude. This sheet therefore shows metallic behaviour even deep in the superconducting state. In this context, it is difficult to ascertain the structure of the large gap from a measurement of the thermal conductivity. Nevertheless, a two-band model with a negligible gap on one Fermi surface and a line node on the other fits the data well, with a universal in-plane conductivity from the line node of magnitude $\kappa_{0a} / T \approx 1.4$ mW / K$^2$ cm (ref. 13). This value is 14 times smaller than in CeIrIn$_5$, but is nevertheless in rough agreement with theoretical expectation for a line node, since $T_c$ (and so presumably the gap maximum) is 6 times larger. (The remaining factor of 2 or so could come from a smaller Fermi velocity on the relevant band.)

The fact that the specific heat[8] and thermal conductivity[7] of CeCoIn$_5$ both exhibit a four-fold anisotropy as a function of magnetic field direction in the basal plane has been interpreted as evidence for *d*-wave symmetry[9], as in the case of CeIrIn$_5$ discussed above. For the same reasons, it is difficult to reliably discriminate between gap anisotropy and Fermi surface anisotropy as the cause of the small four-fold variations detected in these studies, in particular in the presence of strong band dependence of the gap. In the absence of further directional evidence, it is probably fair to say that the location of the line node in the gap structure of CeCoIn$_5$ is still an open question. Nevertheless, it is interesting to contemplate the possibility that the two materials do have different order parameter symmetries, as it could explain the fact that superconductivity in CeIrIn$_5$ appears to reside in a region of the generalized phase diagram of the Ce*M*In$_5$ family (where *M* = Ir, Co, or Rh) which is separate from the region where superconductivity in CeCoIn$_5$ and CeRhIn$_5$ resides[10]. In a scenario of magnetically-mediated superconductivity, a different pairing symmetry would come from a different momentum dependence of the magnetic fluctuations responsible for pairing[1]. The fact that superconductivity coexists with antiferromagnetism in CeCoIn$_5$ and CeRhIn$_5$ but not in CeIrIn$_5$ (ref. 23) may be a manifestation of that difference.

**METHODS**

Single crystals of CeIrIn$_5$ were grown by the self-flux method[14]. Two high-purity crystals were cut so that heat current flowed along the *a*-axis and *c*-axis, respectively. The residual resistivity of these pure samples was $\rho_{0a}$ = 0.2 $\mu\Omega$ cm and $\rho_{0c}$ = 0.50 $\mu\Omega$ cm, respectively. (The difference in $\rho_0$ values is due to mass tensor anisotropy; see ref. 15.) Another two single crystals were grown with a small amount of La impurities (0.1 % of Ce) and cut along the same two directions, with $\rho_{0a}$ = 0.48 $\mu\Omega$ cm and $\rho_{0c}$ = 2.04 $\mu\Omega$ cm, respectively. Addition of La lowered the bulk transition temperature $T_c$ by 10 %, from 0.38 ± 0.02 K in the pure to 0.34 ± 0.02 K in the La-doped samples. Three additional nominally pure *c*-axis samples were measured, with $\rho_{0c}$ = 0.81, 1.18 and 1.65 µΩ cm (see data in Fig. 3). For all samples, $\rho_0$ was obtained by extrapolating the normal-state $\kappa / T$ to $T = 0$ and applying the Wiedemann-Franz law, as described in the caption of Fig. 1 and in ref. 15. The thermal conductivity was measured as described in ref. 15. The relative uncertainty in the absolute value of $\kappa$ between different samples with the same current direction was removed by normalizing their electrical resistivity (measured using the same contacts) to the same value at room temperature, namely $\rho_a(300\ K) = 25.9\ \mu\Omega$ cm and $\rho_c(300\ K) = 52.7\ \mu\Omega$ cm.

**Figure 1 | Effect of impurities on heat transport.**

Thermal conductivity of CeIrIn$_5$ in the superconducting state (magnetic field $H = 0$; full symbols) and normal state ($H = 0.5$ T > $H_{c2} = 0.49$ T; open symbols) of pure (black circles) and La-doped (Ce$_{0.999}$La$_{0.001}$IrIn$_5$; blue squares) samples, plotted as $\kappa / T$ vs $T$ for a heat current parallel ($J \parallel c$; *b*) and perpendicular ($J \parallel a$; *a*) to the *c*-axis of the tetragonal crystal structure. Red arrows show $T_c$. Red lines are an extrapolation of the low-temperature $H = 0$ data to $T = 0$, assuming a linear variation of $\kappa / T$ for the pure $J \parallel a$ data and a $T^2$ variation for the other data. The blue lines are Fermi-liquid fits to the normal-state data (see ref. 15), yielding an extrapolated residual linear term $\kappa_{0N} / T$ from which we obtain the residual resistivity $\rho_0$, via the Wiedemann-Franz law $\kappa_{0N} / T = L_0 / \rho_0$, where $L_0 = 2.45 \times 10^{-8}$ Ω W / K$^2$ (see ref. 15). For $J \parallel a$, the residual linear term $\kappa_0 / T$ is large and independent of impurity content, showing heat conduction in the basal plane to be universal. By contrast, for $J \parallel c$, $\kappa_0 / T$ is negligible in the pure sample and grows to become sizable in the doped sample, showing heat conduction along the *c*-axis to be non-universal.

**Figure 2 | Anisotropy of heat transport.**

**a)** Thermal conductivity of CeIrIn$_5$ as a function of temperature, normalized at $T_c$ = 0.38 K. Data from pure samples are shown for conduction parallel (green squares) and perpendicular (red circles) to the high-symmetry *c*-axis. The low-temperature regime reveals a strong uniaxial anisotropy, with $(\kappa_{0a} / T) / (\kappa_{0c} / T)$ → ∞ as $T$ → 0, consistent with a line node in the basal plane. Lines are guides to the eye. **b)** Sketch of the "hybrid gap", drawn on a single spherical Fermi surface for simplicity, which consists of a line node in the basal plane ($k_z$ = 0; red line) and two linear point nodes ($k_x = k_y$ = 0; green spots) along the *c*-axis (black arrow). **c)** Sketch of the "polar gap", where $\Delta^2 \sim \cos^2\theta$, with $\theta$ the polar angle away from the *c*-axis, which only has a line node in the basal plane ($k_z$ = 0 or $\theta = \pi / 2$; red line).

**Figure 3 | Residual heat conduction vs impurity scattering.**

Residual linear term, $\kappa_0 / T$, in the thermal conductivity of CeIrIn$_5$, as a function of the residual resistivity $\rho_0$ of each sample, a measure of the impurity scattering rate $\Gamma_0$. The value of $\kappa_0 / T$ is obtained by extrapolating $\kappa / T$ vs $T$ data to $T$ = 0 as in Fig. 1. The values of $\rho_0$ are obtained as described in Fig. 1. The error bars on $\kappa_0 / T$ and $\rho_0$ for our seven samples (those with full symbols) reflect the uncertainty in the extrapolation of $\kappa / T$ and $\kappa_N / T$ to $T$ = 0, respectively (see Fig. 1). The relative uncertainty in estimating the geometric factor of these samples is removed by a normalization procedure described in the Methods. The open red circle is the value of $\kappa_{0a} / T$ quoted in ref. 22 for an *a*-axis sample whose $\rho_0$ value we obtain as above by extrapolating the corresponding normal-state data[22]. To the error bar on $\kappa_{0a} / T$ quoted in ref. 22 (± 2 mW / K$^2$ cm), we have added a ± 10 % geometric-factor uncertainty on both $\kappa_{0a}/T$ and $\rho_0$. The in-plane transport (red symbols) shows $\kappa_{0a}/T$ to be constant at ~ 21 mW / K$^2$ cm (red dashed line), independent of $\Gamma_0$, *i.e.* heat conduction in the plane is universal. This proves that there must be a line node in the superconducting gap of CeIrIn$_5$. By contrast, *c*-axis transport (green squares) is not universal, with $\kappa_{0c}/T \to 0$ as $\rho_0 \to 0$. This qualitative difference, combined with the large anisotropy (see Fig. 2), is strong evidence that the line node lies in the basal plane ($k_z = 0$). The solid green line is a linear fit to the *c*-axis data, the dependence expected for a "polar gap" (drawn in Fig. 2c) in the strong impurity scattering limit, where $\kappa_{0c}/T \sim \Gamma_0$ (ref. 18). The dashed green line is a fit to $\kappa_{0c}/T \sim (\rho_0)^{1/2}$, expected theoretically in the same limit for linear point nodes along the *c*-axis (ref. 18), as in the "hybrid gap" drawn in Fig. 2b (ref. 5).

**Acknowledgements** MAT acknowledges his continuous cross-appointment with the Institute of Surface Chemistry, National Academy of Sciences of Ukraine. LT acknowledges support from the Canadian Institute for Advanced Research and funding from NSERC, FQRNT, FCI and a Canada Research Chair. The work was partially carried out at the Brookhaven National Laboratory, which is operated for the U.S. Department of Energy by Brookhaven Science Associates (DE-Ac02-98CH10886).

**Author Information** Correspondence and requests for materials should be addressed to L.T. (louis.taillefer@physique.usherbrooke.ca).

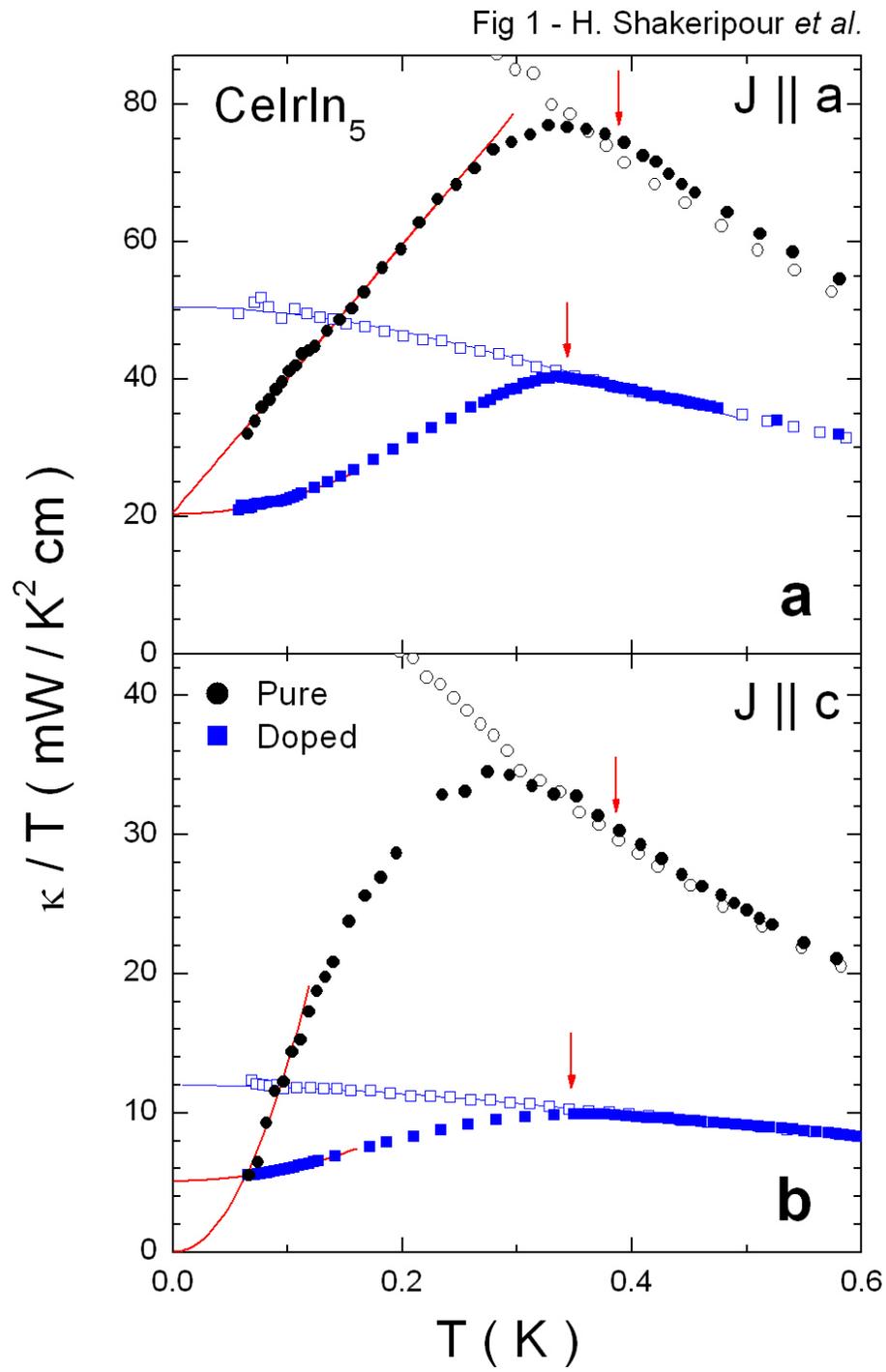

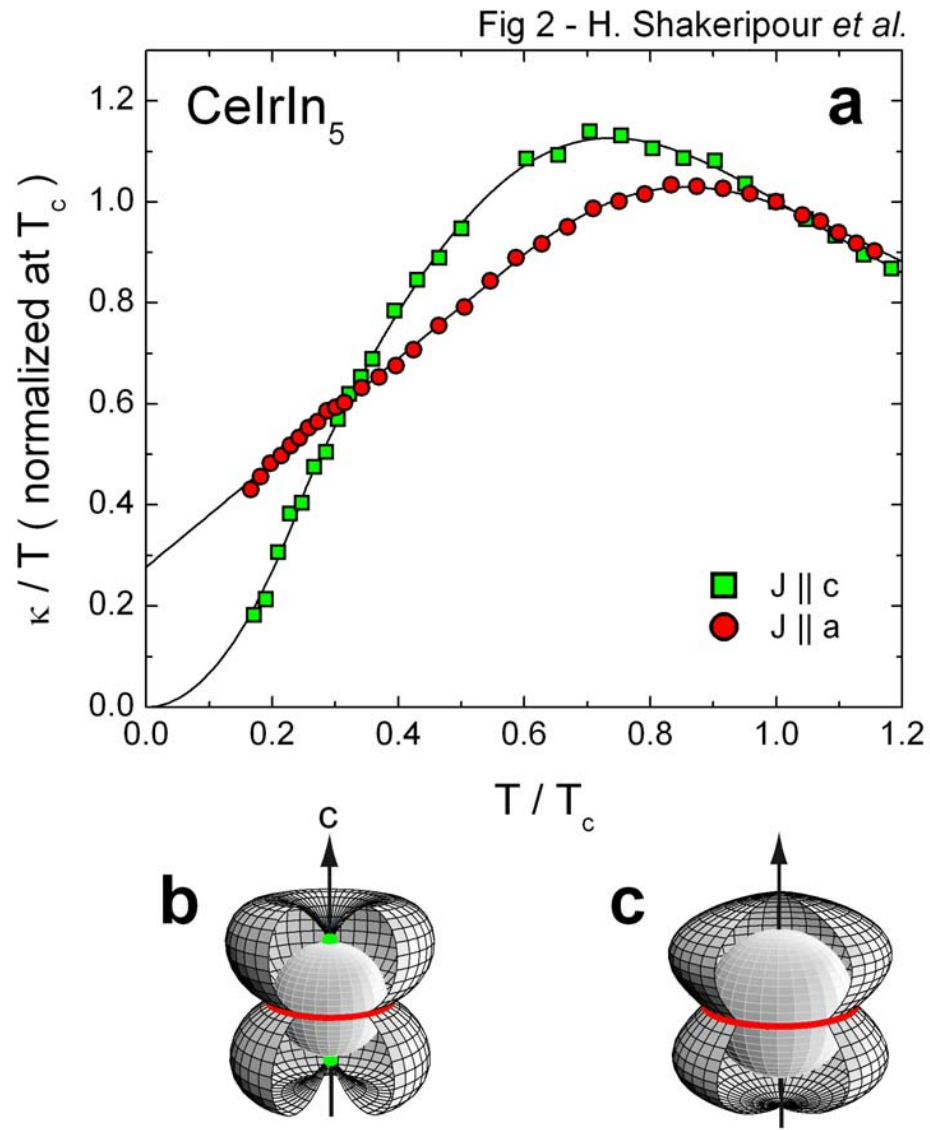

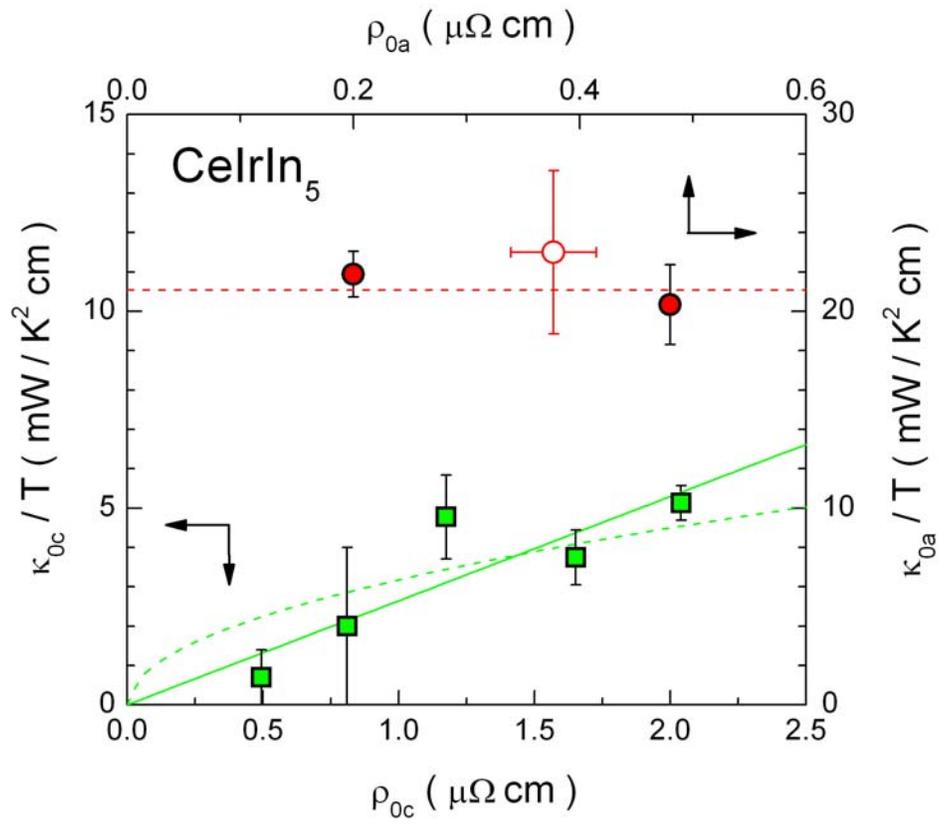

Fig 3 - H. Shakeripour *et al.*